\documentclass[aps,prx,showpacs,notitlepage,floatfix,superscriptaddress,twocolumn, nofootinbib]{revtex4-2}
\usepackage{amssymb}
\usepackage{amsmath}
\usepackage{amsfonts}
\usepackage{bbm}
\usepackage{tikz}
\usetikzlibrary{decorations.pathreplacing}
\usepackage{graphicx}
\usepackage{braket}
\usepackage[colorlinks,linkcolor=blue,anchorcolor=blue,citecolor=blue,urlcolor=blue]{hyperref}
\usepackage[titletoc]{appendix}

\def\S{\mathcal{S}}
\def\tr{\text{tr}}
\def\L{\mathbb{L}} 
\def\H{\mathbb{H}}

\graphicspath{{Figures/}}

\begin{document}
\title{Non-Abelian operator size distribution in charge-conserving many-body systems}
\begin{abstract} 
We show that operator dynamics in U(1) symmetric systems are constrained by two independent conserved charges and construct a non-Abelian operator size basis that respects both, enabling a symmetry-resolved characterization of operator growth. The non-Abelian operator size depends on the operator’s nonlocal structure and is organized by an SU(2) algebra. Operators associated with large total angular momentum are relatively simple, while those with small angular momentum are more complex. Operator growth is thus characterized by a reduction in angular momentum and can be probed using out-of-time-ordered correlators. Using the charge-conserving Brownian Sachdev-Ye-Kitaev model, we derive an exact classical master equation that governs the size distribution, the distribution of an operator expanded in this basis, for arbitrary system sizes. The resulting dynamics reveal that the size distribution follows a chi-squared form, with the two conserved charges jointly determining the overall time scale and the shape of the distribution. In particular, single-particle operators retain a divergent peak at large angular momentum throughout the time evolution.
\end{abstract}
\author{Mina Tarakemeh}
\affiliation{Department of Physics \& Astronomy, Texas A\&M University, College Station, Texas 77843, USA}
\author{Shenglong Xu}
\email{slxu@tamu.edu}
\affiliation{Department of Physics \& Astronomy, Texas A\&M University, College Station, Texas 77843, USA}

\maketitle
  
In generic quantum many-body systems, a simple operator in the Heisenberg picture spreads and becomes increasingly complex and nonlocal, a universal behavior that emerges despite system-specific details~\cite{nahum2018operator, von2018operator, parker2019universal}.  This growth underlies key non-equilibrium phenomena such as thermalization~\cite{rigol2008thermalization, polkovnikov2011colloquium, kaufman2016quantum}, quantum transport~\cite{kim2015slowest,rakovszky2022dissipationassisted}, and information scrambling~\cite{lewis-swan2019unifying, xu2022scrambling}.
The standard approach to quantify the operator growth is to identify the operator size basis, a complete set of operators in which each basis operator $\hat{\S}_i$ has a definite ``size" $s$ characterizing its complexity and $i$ indexes the degeneracy. Expanding a Heisenberg operator $O(t)$ in this basis, one can define the operator size distribution $P(s,t)$ by aggregating all the basis operators of size $s$.
\begin{equation}
O(t) = \sum_{s,i} a_i(s,t) \hat{\S}_i, \quad P(s,t) = \sum_i |a_i(s,t)|^2.
\end{equation}
 Operator growth corresponds to a shift of weight in $P(s,t)$ from small to large sizes. In systems without conserved quantities, the size is based on the support of the operators. In qubit systems, the operator size basis consists of Pauli strings, products of local Pauli operators and identities, with size given by the number of non-identity Pauli operators~\cite{nahum2018operator, xu2019locality}. Similarly, in Majorana systems~\cite{roberts2018operator, zhang2023operator}, the basis is formed by Majorana strings. The growth of operators and the time evolution of operator size distributions have been studied extensively~\cite{zhou2019operator, qi2019quantum, roberts2018operator, lucas2020nonperturbative, zhang2023operator, yao2024notessolvable, xu2025dynamics}, and are well captured by classical Markovian dynamics~\cite{nahum2018operator, xu2019locality, sunderhauf2019quantum, You2018, zhou2019operator, kuo2020markovientanglement,xu2025dynamics}.  The averaged size is related to the out-of-time-ordered correlator~\cite{larkin1969quasiclassical, xu2022scrambling}, which has been measured in various quantum devices~\cite{mi2021information, landsman2019verified, garttner2017measuring,braumuller2021probing, wang2021verifying,sanchez2020perturbation}. It was recently proposed to directly probe the entire distribution in quench experiments~\cite{qi2019measuring}. 
%The expectation value of the operator size, $ \braket{\mathbb{S}} = \tr(O^\dagger \mathbb{S}(O))$, corresponds to the first moment of $P(s,t)$. 

Operator growth in systems with conservation laws exhibits richer phenomena~\cite{rakovszky2018diffusive, khemani2018operator, pai2019localization, cheng2021scrambling, hunter-jones2018operator,feldmeier2021critically, kudler-flam2022information, cheng2025hydrodynamic}. 
%Conservation laws constrain operator dynamics to specific symmetry sectors, restricting their ability to spread freely. 
Defining operator size in a symmetry-respecting way becomes nontrivial: an operator confined to a symmetry sector is often nonlocal, while a local operator with finite support may overlap multiple symmetry sectors. Previous work~\cite{chen2020manybody,zhang2023information} has circumvented this issue by working in the grand canonical ensemble, which relaxes the strict symmetry constraints. However, this approach does not apply to operators restricted to fixed symmetry sectors. 
This tension between symmetry and locality~\cite{marvian2022restrictions, lastres2024nonuniversality} demands a new understanding of operator size in systems with conservation laws.
While operator growth can always be characterized by Krylov complexity~\cite{parker2019universal, caputa2025growth}, the Krylov basis depends on both the initial operator and the details of underlying dynamics and lacks a transparent physical interpretation.

%Operator growth in systems with conservation laws display...

%However, defining operator size and its distribution in systems with symmetry can be challenging. For example, in fermionic systems with charge conservation, a natural choice for the basis operator string is ${c^\dagger, c, I, n}$. Previous work defined the size of the operator string by assigning weight 1 to $c$ and $c^\dagger$, and weight 2 to $n$. This approach was successfully applied to characterize the growth of local operators such as $c(t)$ and $n(t).$ However, this scheme is incompatible with the U(1) symmetry and leads to problems for operator dynamics within a fixed symmetry sector. For instance, the operator $n_1 n_2 \cdots n_N$ projects onto the maximally filled Hilbert space and should have a size of 0, but the previous scheme assigns it a size of $2N$.

\begin{figure}
    \centering
    \includegraphics[width=\columnwidth]{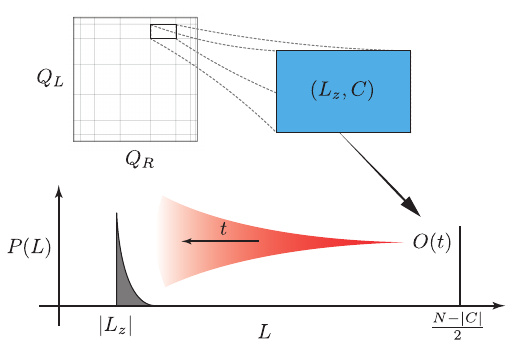}
    \caption{Sketch of the symmetry resolved operator growth in U(1) symmetric systems within each charge sector.  Operator growth is captured by the reduction of total angular momentum from the near maximal value $N/2$ down to $|L_z|$. }
    \label{fig:cover}
\end{figure}

In this work, we present a new definition of operator size fully compatible with symmetries in many-body systems with charge conservation. In this case, the Heisenberg time evolution is constrained by two independent conserved charges, and we construct an SU(2) operator algebra in which the onsite operators $n=c^\dagger c$ and $\bar{n} = I - n$ form a spin-1/2 doublet, while $c$ and $c^\dagger$ are treated as singlets. The operator size is identified with the total angular momentum  $L$ of the SU(2) algebra, which commutes with both conserved quantities. Operators with maximal $L$ correspond to projectors onto individual charge sectors. Large angular momentum is associated with simple operators, and operator growth thus manifests as the reduction of angular momentum down to the fixed $L_z$ set by the charges~(Fig.~\ref{fig:cover}).

Furthermore, we derive an exact classical master equation governing the evolution of the SU(2) size distribution for the Brownian complex Sachdev-Ye-Kitaev~(SYK) model, by fully leveraging the model's symmetries~\cite{agarwal2021emergent}.
The equation's dimension scales linearly with system size, enabling efficient large-scale numerical simulations. We further provide an analytical solution that reveals that the size distribution follows a chi-squared distribution of a dynamical variable.

\textit{Symmetry in operator dynamics --} We focus on U(1) symmetric many-body systems of $N$ complex fermions described by the Hamiltonian in the form of $\sum J_{ij} c^\dagger_i c_j + J_{ijkl} c^\dagger_i c^\dagger_j c_k c_l + \cdots + h.c. $
where $c^\dagger_i / c_i$ is the creation/annihilation operator on site $i$ and $n_i = c^\dagger_i c_i$ is the number operator. 
The conserved quantity is the total charge $Q = \sum_i n_i$, which ranges from $0$ to $N$. Charge conservation decomposes the operator matrix into $(N+1)^2$ charge sectors, and an operator initialized in a given sector remains confined to it~(Fig.~\ref{fig:cover}). A key implication of U(1) symmetry in the Heisenberg picture is therefore the emergence of \textit{two} independent conserved charges~\cite{khemani2018operator}, denoted $Q_L$ and $Q_R$, which label these sectors. These charges can be measured by
\begin{equation}
    Q_L= \text{tr} (O^\dagger(t) Q O(t)),\quad  Q_R=\text{tr} (O^\dagger (t)  O(t) Q ),
\end{equation}
where the operator $O(t)$ is normalized such that $\tr(O^\dagger(t) O(t)) = 1$. 
We also introduce $(L_z, C)$, a linear independent combination of $(Q_L, Q_R)$
\begin{equation}
    L_z = (Q_L+Q_R-N)/2, \quad C = Q_L - Q_R.
\end{equation}
Here, $L_z$ ranges from $-N/2$ to $N/2$ while $C$ ranges from $-N$ to $N$. 
The diagonal sectors with $Q_L = Q_R$~($C=0$) correspond to operators that preserve the total charge, such as $n_i$ and $c_i^\dagger c_j$. In contrast, off-diagonal sectors with $Q_L \neq Q_R$~($C\neq0$) correspond to operators that change the total charge, such as $c_i$ and $c_i^\dagger$.

A general operator may span multiple charge sectors, but its dynamics can be analyzed independently within each sector, and the full evolution is obtained by summing over all sectors.
Studying charge-resolved operator growth therefore requires an operator-size basis with two key properties:
\begin{enumerate}
    \item The basis respects both $L_z$ and $C$, such that each basis operator resides entirely within one of the $(N+1)^2$ charge sectors.
    \item The operator size captures the complexity of operators within each charge sector. For instance, the global identity operator and the projection operators onto fixed charge sectors, which are static under time evolution, should be considered simple operators with size zero.
\end{enumerate}

The standard definition of operator size, based on spatial support, does not extend naturally to the symmetry-resolved setting. An operator with finite support typically spans many $L_z$ sectors, whereas an operator confined to a single charge sector is generally highly nonlocal. For example, the global identity operator has zero spatial support yet decomposes into a sum of projection operators that live entirely in diagonal charge sectors and act nonlocally across all sites.

% Previous works~\cite{chen2020many, zhang2023information} adopted a product operator basis from the local orthonormal set $\{I, c^\dagger, c,, 2n - I\}_i$ which are assigned weight ${0, 1, 1, 2}$,  and the size of a basis operator is defined as the sum of the weights of $N$ local operators. However, a basis operator in this basis typically spans multiple $L_z$ sectors, and the only operator with size zero is the global identity operator.

\textit{Non-abelian operator size -- } We introduce an SU(2) operator-size basis that satisfies the two desired properties, enabling a fully symmetry-resolved characterization of operator growth. We begin with product operator strings constructed from a local orthogonal operator set $\{ c^\dagger, c, n, \bar{n} \}_i$, where $\bar{n}_i = I - n_i$ projects onto the empty site. Each operator string has well-defined values of $Q_L$ and $Q_R$, determined by simple counting: $L_z$ measures half the difference between the numbers of $n$ and $\bar{n}$ operators, and $C$ counts the difference between the numbers of $c^\dagger$ and $c$ operators.

%We note that this basis is fully compatible with the symmetry at the cost of locality since the identity operator is \textit{not} a basis operator. 

We construct an SU(2) algebra of an operator by treating $n$ and $\bar n$ as the spin-1/2 doublet, and $c$ and $c^\dagger$ as the singlet,
\begin{equation}
n \rightarrow \ket{\uparrow}, \ \  \bar{n} \rightarrow \ket{\downarrow}, \ \ \{c^\dagger, c \}\rightarrow \ket{0}.
\end{equation}
The angular momentum super operators $\vec{\mathbb{ L}} =\sum _i \vec {\mathbb{L}}_i$ are a sum of local super operators 
\begin{equation}
\label{eq:SU(2)}
    \L^+_i(O) = c^{\dagger}_i O  c_i, \  \L^-_i(O) = c_i O  c_i^\dagger,  \ \L^z_i(O) =\frac{1}{2}(n_i O - O \bar n_i).
\end{equation}
It is straightforward to verify the SU(2) commutation relation and that $\vec {\L}_i$ annihilates $c$ and $c^\dagger$ and acts on $n$ and $\bar n$  as if they are spin-1/2 doublets. These super operators conserve $C$, and $\L_z$ directly measures $L_z$. The total angular momentum operator $\L^2 = (\L^+ \L^- + \L^- \L^+)/2 + (\L^z)^2$ commutes with both $\L_z$ and $C$. We obtain the SU(2) operator size basis as the eigen operators of $\L^2$. Each basis operator $\mathcal{L}^{L_z,C}_i$ is labeled by the quantum numbers $(L, L_z, C)$, where $L(L+1)$ is the eigenvalue of $\L^2$, and $|L_z|\leq L\leq (N-|C|)/2$. The degeneracy of each $(L, L_z, C)$ sector is given by counting the number of ways to distribute $n$ and $\bar n$ operators to form total angular momentum $L$ with fixed $L_z$, and distributing $c$ and $c^\dagger$ operators to satisfy the fixed $C$, leading to~(see derivation in the appendix)
\begin{equation}\label{eq:degeneracy}
\begin{aligned}
   D^{L_z, C}(L) =
    \frac{1+2L}{N+1}\begin{pmatrix} N+1\\ \frac{N-C}{2}-L\end{pmatrix}\begin{pmatrix} N+1\\ \frac{N+C}{2}-L\end{pmatrix}
\end{aligned}
\end{equation}
A complete list of the 16 two-site basis operators is provided in the appendix.

The SU(2) operator size basis satisfies the two desired properties for symmetry-resolved operator growth. First, by construction, each basis operator carries good quantum numbers $L_z$ and $C$. Second, the basis naturally captures the complexity of operators through their angular momentum. Operators with large angular momentum are relatively simple, while those with small angular momentum are more complex and have larger degeneracy in Eq.~\eqref{eq:degeneracy}. The global identity operator $I = \prod(n_i + \bar n_i)$ is mapped to the fully polarized state in the $x$ direction and thus has the largest angular momentum $L = N/2$. The $(N+1)$ projection operators onto fixed charge sectors also have the largest angular momentum and correspond to different $L_z$ components, reflecting their simple structure and static nature. Furthermore, remarkably, any operator with a finite support has a well-defined angular momentum $L$ with a finite offset $s = N/2-L$ from the maximal value $N/2$ in the thermodynamic limit. 
In the appendix, we show that for a local operator acting on $m_c$ sites with $c$ or $c^\dagger$ and $m_n$ sites with $2n -I$, $s = m_c/2 + m_n$. Thus, for local operators, the SU(2) operator size reduces to the previous definition based on counting local operators, where $2n-I$ is double weighted compared to $c$ and $c^\dagger$~\cite{chen2020many, zhang2023information}. 

In our framework, a simple operator is associated with a large angular momentum, and operator growth is characterized by the \textit{reduction} of angular momentum, as illustrated in Fig.~\ref{fig:cover}.  Then it is natural to define the size as $s = N/2 - L$,  which increases over time. However, we remark that $\braket{\L^2}$ is an experimentally accessible observable. From Eq.~\eqref{eq:SU(2)}, we have
% From the $\L$ operators in Eq.~\eqref{eq:SU(2)}, we get
\begin{equation}\label{eq:lsquare}
\begin{aligned}
    \braket{\L^2(t)} =\sum\limits_{ij}\tr \left(O^\dagger(t) c_i^\dagger c_j O(t) c_j^\dagger c_i \right) + \braket{L_z(L_z-1)}.
\end{aligned}
\end{equation}
The first term is an OTOC, which has been experimentally measured, and the second term remains a constant throughout the time evolution. 
We emphasize that this is valid even for operators spanning multiple charge sectors.

% In the late time, ergodicity implies that every basis operator is equally probable and the size distribution is proportional to the degeneracy $D$ in Eq.~\eqref{eq:degeneracy}. 

\textit{Symmetry resolved Operator Size Dynamics -- } 
Now we study the dynamics of the SU(2) size distribution, i.e., how the distribution of an initial operator in this size basis converges to the steady distribution as a function of time, which offers more detailed information than the average size $\braket{\L^2}$.
Consider a Heisenberg operator expanded in the basis, $O(t) = \sum _{L,i}^{L_z, C}a_i^{L_z, C}(L,t) \mathcal{L}_i^{L_z,C}$. We define the charge resolved operator size distribution $P^{L_z,C}(L,t) = \sum_i |a_i^{L_z, C}(L,t)|^2$ by aggregating basis operators in the same charge sector carrying the same $L$. 

We employ the Brownian complex SYK model~\cite{Lashkari2012, saad2018semiclassical} with charge conservation to study the operator size dynamics. The Hamiltonian is given by
\begin{equation}
\label{eq:brownian}
H = \sum_{ijkl} J_{ijkl}(t) c_i^\dagger c_j^\dagger c_k c_l +h.c.    
\end{equation}
The couplings are Gaussian white noise obeying
\begin{equation}
   \overline{J_A(t)}=0,\ \ \overline{J_A(t) J^*_{A'}(t')}
   =\delta_{AA'}\delta(t-t')J/(2N^3)
\end{equation} 
where $A$ represents the grouped index $ijkl$ and the energy unit $J$ is set to 1. This model is the continuous-time limit of a random quantum circuit~\cite{fisher2022random} composed of few-body, charge-conserving gates.

The Brownian SYK model is widely used to study operator dynamics, entanglement growth, and complexity in the large-$N$ limit~\cite{saad2018semiclassical, stanford2022subleading, zhang2023information, zhang2023operator, jian2021note, jian2022linear, tiutiakina2023frame}. Here we instead employ a noise-averaging approach at finite $N$ that remarkably yields an \textit{exact} classical master equation for the SU(2) operator-size dynamics. Averaging over the random couplings maps real-time dynamics to an imaginary-time evolution governed by an effective super Hamiltonian acting on four replicas~\cite{sunderhauf2019quantum, agarwal2021emergent, jian2022linear}. The low-lying modes of the effective Hamiltonian control the long-time behavior of observables and reflect the symmetries of the underlying model~\cite{singh2021subdiffusion, moudgalya2023symmetries, vardhan2024entanglement, ogunnaike2023unifying}. Interestingly, the effective Hamiltonian also exhibits emergent symmetries absent in single realizations~\cite{bao2021symmetry, bernard2022dynamics, agarwal2021emergent, swann2023spacetime}: it is invariant under replica permutations and depends only on uniform fermion bilinears generating an SU(4) algebra. Although not fully SU(4)-symmetric, the effective Hamiltonian commutes with the SU(4) Casimir, and the Hilbert space is fragmented into exponentially many sectors 
labeled by SU(4) irreducible representations~(irrep) and conserved charges. The dimension of each block scales linearly in $N$~\cite{bernard2022dynamics, agarwal2021emergent}, and the Hamiltonian can be explicitly calculated using SU(4) operators in that irrep~\cite{alex2011numerical}. The SU(2) operator size basis we construct spans one such symmetry sector, where the emergent Hamiltonian reduces to a classical master equation~(see details and full derivation in the Appendix).

The master equation, exact for arbitrary $N$ and all charge sectors labeled by $(L_z, C)$, reads
\begin{equation}\label{eq:masterEq}
\begin{aligned}
% \partial_t P(L)  = \sum_{L'} (\delta_{L, L'-1} -\delta_{L,L'})\gamma^-_{L'} + (\delta_{L, L'+1} -\delta_{L,L'})\gamma^+_{L'} P(L')
%     \\
    &\partial_t P(L)=
    \\&\quad\gamma^-_{L+1} P(L+1)-\gamma^-_L P(L)+
    \gamma^+_{L-1} P^{}(L-1)-\gamma^+_L P(L)
\end{aligned}
\end{equation}
where $\gamma^\pm$ are
\begin{equation}\label{eq:rates}
\begin{aligned}
    \gamma^-_L&=\frac{( (N+2L+2)^2-C^2)(N^2-C^2-4L^2)(L^2-L_z^2)}{8L(2L+1)N^3} \\
    \gamma^+_{L-1} &=\frac{( (N-2L+2)^2-C^2)(N^2-C^2-4L^2)(L^2-L_z^2)}{8L(2L-1)N^3} 
\end{aligned}
\end{equation}
representing the transition rates. The range of $L$ is $|L_z|\leq L\leq (N-|C|)/2$.
We have verified that the distribution in Eq.~\eqref{eq:degeneracy} is the stable static solution, indicating late-time ergodicity. In addition, for diagonal sectors where $C=0$, there is an unstable static solution localized at $L=N/2$ representing the projector operator. The benchmarking against direct numerical simulations of Eq.~\eqref{eq:brownian} for small system sizes shows perfect agreement~(Appendix).
\begin{figure}
    \centering
    \includegraphics[width=\columnwidth]{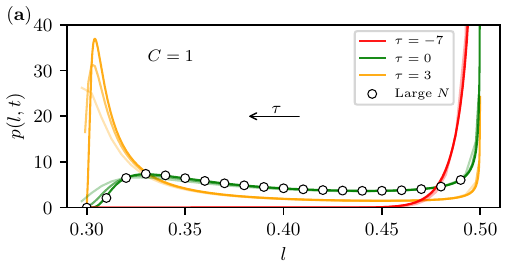}
    \includegraphics[width=\columnwidth]{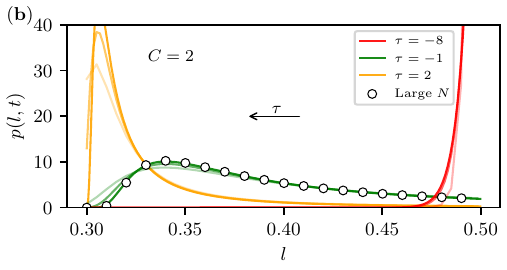}
    \includegraphics[width=\columnwidth]{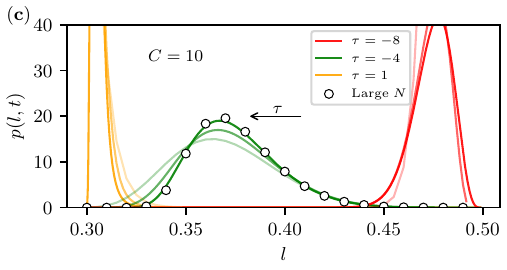}
    \caption{Numerical simulations of the master equation for system sizes $N = 200$--$4000$, with darker curves indicating larger $N$. The shifted time $\tau = t - t^*$ is held fixed as $N$ increases. The large $N$ distributions exhibit excellent agreement with the chi-squared form in Eq.~\eqref{eq:chi}.  Panels (a), (b), and (c) show results for $C = 1$, $2$, and $10$, respectively. All simulations use $l_z = 0.3$ and $L_0 = (N - C)/2$.
}
    \label{fig:simulation}
\end{figure}

The exact master equation scales linearly with $N$, enabling efficient simulation for large systems and avoiding the exponentially large Hilbert space dimension before the random average. We consider a simple local operator with a finite initial size, $s_0 = N/2 - L_0$. From the range of $L$, $s_0> |C|/2$, and therefore $C$ is also finite. However, such an initial operator typically spans multiple $L_z$ sectors, and we consider the limit that $l_z = L_z/N$ is fixed as $N$ increases. 

Owing to the finite initial size $s_0$, the operator distribution evolves through two regimes: an early-time regime with $s = \mathcal{O}(1)$, where it remains localized near the initial point, and an intermediate/late-time regime with $s = \mathcal{O}(N)$, where it escapes from the initial point.
In the early time regime, taking the large $N$ limit with fixed $C$ and $l_z$ leads to a master equation describing a pure birth process:
\begin{equation}\label{eq:early-time}
    \partial_t P(s) = \lambda \left[(s - 1)P(s - 1) - sP(s)\right], \quad s=N/2 - L.
\end{equation}
where $\lambda = 1 - 4l_z^2$, sets a charge-dependent time scale.  As time increases, the discrete distribution spreads and can be approximated by a continuous distribution function $p(l,t)$, whose dynamics is governed by the Fokker-Planck equation
\begin{equation}\label{eq:fokker-planck}
    \partial_t p(l) = -\partial_l \left( v(l)\, p(l) \right)  + \frac{1}{N} \partial_l^2 \left( d(l)\, p(l,t) \right),
\end{equation}
where $v(l)=\frac{2}{l}(l^2-l_z^2){(l^2 - l_z^2 -\lambda/4 )}$ is the drift coefficient. The second term is suppressed by a large $N$ and neglected.  We solve both equations and match the distribution at the crossover time to obtain the full-time distribution (see appendix and \cite{xu2025dynamics} for details). The solution is a chi-squared distribution $f_{s_0}(\xi)$ of a new dynamical variable \( \xi(l, t) \):
\begin{equation}\label{eq:chi}
    p(l,t) = \frac{d \xi}{dl} f_{s_0}(\xi) = \frac{d\xi}{dl} \cdot \frac{1}{\Gamma(s_0)} e^{-\xi} \xi^{s_0 - 1},
\end{equation}
The degree of the chi-squared distribution is set by the initial size $s_0$. The new variable $\xi$, solely determined by the velocity $v(l)$ is
\begin{equation}
     \xi = \frac{e^{(t^* - t)\lambda} \, \lambda(1 - 4l^2)}{16\left(l^2 - l_z^2\right)}, \quad t^* = \frac{\log N}{\lambda}. 
\end{equation}
where $t^*$ is interpreted as the scrambling time. This distribution is valid when $t-t^*$, the difference between the physical time and the scrambling time, is finite compared with $N$.  
We numerically solve the master equation in Eq.~\eqref{eq:masterEq} to obtain the full time-dependent operator size dynamics for various values of $C$  and system sizes up to $N = 4000$, and find excellent agreement with the chi-squared form in Eq.~\eqref{eq:chi} as shown in Fig.~\ref{fig:simulation}.

The analytical solution highlights the distinct roles of $L_z$ and $C$ in symmetry-resolved dynamics. The scaled quantity $l_z$ enters the variable $\xi$ by setting the timescale $\lambda = 1 - 4l_z^2$ and the equilibrium value $l_z^2$, but does not affect the chi-square distribution $f_{s_0}$.  Eq.~\eqref{eq:chi} leads to the average operator angular momentum \( \langle l^2 \rangle \):
\begin{equation}\label{eq:lsquare_smallC}
    \langle l^2\rangle= \frac{1}{4}\left(1-e^x s_0 \lambda E_{1+s_0}(x)\right), \quad x=\frac{1}{4}e^{\lambda (t^*-t)}\lambda
    % \frac{1}{16}e^{\frac{e^{-\tau \lambda}}{4}-\tau \lambda}(1-4l_z^2)E_{s_0}(\frac{e^{-\tau \lambda}}{4})
\end{equation}
\( E_{1 + s_0}(x) \) denotes the exponential integral function. In the early-time regime $t\ll t^*$, the average size grows exponentially
$
    \langle l^2 \rangle -1/4 \approx - s_0 \, e^{\lambda \tau},
$
indicating fast scrambling with a charge dependent Lyapunov exponent $\lambda$. 

In contrast, $C$ does not directly enter the analytical form but imposes a lower bound on $s_0$, namely $s_0 \geq C/2$, thereby determining the shape of the chi-square distribution. Notably, when $s_0 = 1/2$, which only occurs for $C=1$ corresponding to single particle operators $c_i^\dagger$ or $c_i$, the size distribution exhibits a square-root divergence at $l=1/2$ at all time across different $l_z$. This divergence is visible in the numerical data in Fig.~\ref{fig:simulation}(a) for $C = 1$, but absent in Fig.~\ref{fig:simulation}(b) and (c) for $C = 2$ and higher. This suggests the single-particle Heisenberg operator still retains certain memory and grows more slowly compared with other operators. 

As $C$ increases, the initial size $s_0$ also increases, leading to a higher-degree chi-squared distribution, which approaches a Gaussian form according to the central limit theorem, resulting in a narrower size distribution~(Fig.~\ref{fig:simulation}(c)). As $C$ further increases and scales with $N$, the initial size $s_0 = \mathcal{O} (N)$, and the size dynamics skips the early time regime entirely, governed solely by the Fokker-Planck equation in Eq.~\eqref{eq:fokker-planck}. The size distribution in this case is described by a delta function centered at the average. The drift coefficient stays the same form but with $\lambda = 1- c^2 - 4l_z^2$ where $c= C/N$.   Solving the corresponding characteristic equation $d_t l = v(l)$, we find that
\begin{equation}
\label{eq:lsquare_largeC}
    l^2(t) = \frac{\lambda}{4 + 4e^{\lambda (t - t_0)}} + l_z^2, \quad \lambda = 1 - c^2 - 4l_z^2
\end{equation}
where $t_0$ is determined by the initial condition. For a generic charge sector, the scaled charges $(l_z, c)$ jointly set the time scale $\lambda$.

\textit{Discussion and Outlook -- } 
In this work, we show that U(1) symmetric operator dynamics respects two conserved quantities $(L_z, C)$ and formulate operator growth by defining operator size as the total angular momentum in an operator SU(2) algebra, fully compatible with the system's conserved charges. For operators with a finite support, the new definition reduces to the previous one based on counting local operators. We derive an exact classical master equation governing the operator size dynamics in the Brownian complex SYK model with charge conservation, valid for arbitrary system size and all charge sectors. The equation's linear scaling with system size enables efficient numerical simulations, and we provide an analytical solution revealing a chi-squared distribution of a dynamical variable $\xi(l,t)$. The conserved charges play distinct roles: $L_z$ sets the equilibrium value, while $C$ determines the distribution's shape, and they together set the time scale of operator growth.

This work reveals the intrinsic relation between operator size and the symmetry of the system. An interesting question is how to define operator size in the presence of the other symmetries,  a direction we leave for future study. It also establishes the charge-conserving Brownian SYK model as a tractable finite-$N$ strongly-interacting model for exploring a variety of other dynamical phenomena constrained by conserved charges.

\textit{Acknowledgment} -- S. Xu thanks Brian Swingle,  Xiaoliang Qi, Tianci Zhou, and Yingfei Gu for useful discussions. 
This work is supported by the National Science Foundation under grant No. DMR-2443462. S. Xu also acknowledges the Google Research Scholar Program and the advanced computing resources provided by Texas A\&M High Performance Research Computing.

\textit{Data Availability} -- The code and resulting data for the numerical simulation of the master equation and small $N$ benchmark are available at Zenodo~(\url{https://doi.org/10.5281/zenodo.17663044}).

\bibliography{reference}
\onecolumngrid
\appendix

\section{SU(2) operator size basis for system with $U(1)$ symmetry}
In this section, we derive the degeneracy factor for the SU(2) operator basis with a fixed $L$, $L_z$ and $C$.  As discussed in the main text, this basis can be mapped onto an SU(2) spin language, where $n$ and $\bar{n}$ are treated as spin-$1/2$ doublet, while $c$ and $c^\dagger$ form spin singlets with angular momentum $L=0$. In this representation, each operator string is an eigenoperator of $ \mathbb{L}^2$ with eigenvalue $L(L+1)$, characterized by fixed values of $C= Q_L - Q_R$ and $L_z = (Q_L + Q_R-N)/2$.
To enumerate all basis elements with given $L$, $L_z$, and $C$, we consider operator strings in which $n$ or $\bar{n}$ acts on $k$ number of sites, while the remaining $N - k$ sites are occupied by $c$ and $c^\dagger$. The number of spin-$1/2$ configurations on these $k$ sites with total spin $L$ and $L_z$ is given by:
\begin{equation}
    d_k(L) = \binom{k}{k/2 - L} - \binom{k}{k/2 - L - 1}.
\end{equation}
Since $c$ and $c^\dagger$ are spin singlets, they do not contribute to the total angular momentum $L$. However, they do affect the conserved charge $C$ and occupy sites within the operator string. Taking into account the total number of sites $N$ and the charge conservation constraint, we have
\begin{equation}
    N= k + \#c + \#c^\dagger , \quad C = \#c^\dagger - \#c ,
\end{equation}
The number of distinct ways to distribute $c$ and $c^\dagger$ operators among the remaining $N - k$ sites is therefore given by
\begin{equation}
    \binom{N - k}{\#c} = \binom{N - k}{\frac{N - C - k}{2}}.
\end{equation}
The total number of operator basis states with fixed \( L \), \( L_z \), and \( C \) is obtained by summing over all allowed values of \( k \):
\begin{equation}
    D^{L_z, C}(L) = \sum_k d_k(L) \binom{N - k}{\frac{N - C - k}{2}} \binom{N}{k}.
\end{equation}
We evaluate this sum analytically and obtain the following closed-form expression:
\begin{equation}
\begin{aligned}
   D^{L_z, C}(L) =
    \frac{1+2L}{N+1}\begin{pmatrix} N+1\\ \frac{N-C}{2}-L\end{pmatrix}\begin{pmatrix} N+1\\ \frac{N+C}{2}-L\end{pmatrix}.
\end{aligned}
\end{equation}

\section{Operator basis and charge sector decomposition for two sites}
We consider the full operator basis constructed from $\{c, c^\dagger, n, \bar{n}\}$ on $N$ lattice sites, which yields $4^N$ distinct many-body operators. Due to the global $U(1)$ symmetry, the operator space decomposes into $(N+1)^2$ charge sectors labeled by the left and right charges $(Q_L, Q_R)$, or equivalently by $C = Q_L - Q_R$ and angular momentum component $L_z = (Q_L + Q_R - N)/2$. As a concrete example, we present all 16 independent operators for $N = 2$ in Fig~\ref{fig:grid}. Each cell corresponds to a fixed charge sector and lists the full set of linearly independent operators in that block. Alongside each operator, we indicate its total spin $L$, reflecting the underlying SU(2) structure. Under Heisenberg time evolution, an operator remains confined within its original $(Q_L, Q_R)$ block due to charge conservation. However, it dynamically evolves among different $L$ subspaces within that block.

\begin{figure}[htbp]
\centering
\begin{tikzpicture}[every node/.style={font=\small,align=center}]

% -- Include your cell size definitions and drawing code here --% Cell sizes
\def\smallwidth{1.75cm}
\def\smallheight{1.25cm}
\def\bigwidth{4cm}
\def\bigheight{2.5cm}

\node at (-0.5, \smallheight + 0.5*\bigheight) {$Q_L$};

% Label on the top (Q_R), horizontally centered
\node at (\smallwidth + 0.5*\bigwidth, \smallheight + \bigheight + \smallheight + 10) {$Q_R$};

% Coordinates for each cell (i,j), with (1,1) being the center cell
% Top row
\draw (0, \smallheight + \bigheight) rectangle ++(\smallwidth, \smallheight);  % (0,2)
\draw (\smallwidth, \smallheight + \bigheight) rectangle ++(\bigwidth, \smallheight); % (1,2)
\draw (\smallwidth + \bigwidth, \smallheight + \bigheight) rectangle ++(\smallwidth, \smallheight); % (2,2)

% Middle row
\draw (0, \smallheight) rectangle ++(\smallwidth, \bigheight); % (0,1)
\draw (\smallwidth, \smallheight) rectangle ++(\bigwidth, \bigheight); % center (1,1)
\draw (\smallwidth + \bigwidth, \smallheight) rectangle ++(\smallwidth, \bigheight); % (2,1)

% Bottom row
\draw (0, 0) rectangle ++(\smallwidth, \smallheight); % (0,0)
\draw (\smallwidth, 0) rectangle ++(\bigwidth, \smallheight); % (1,0)
\draw (\smallwidth + \bigwidth, 0) rectangle ++(\smallwidth, \smallheight); % (2,0)

% Nodes in each cell
\node[align=center] at (0.5*\smallwidth, \smallheight + \bigheight + 0.5*\smallheight) {$\bar{n}_1 \bar{n}_2$\\
$ (L = 1)$}; % (0,2)

\node[align=center] at (\smallwidth + 0.5*\bigwidth, \smallheight + \bigheight + 0.5*\smallheight) {$\bar{n}_1 c_2 \quad (L = \frac{1}{2})$ \\
$c_1 \bar{n}_2 \quad (L = \frac{1}{2})$}; % (1,2)

\node[align=center] at (\smallwidth + \bigwidth + 0.5*\smallwidth, \smallheight + \bigheight + 0.5*\smallheight) {$c_1 c_2$ \\
$(L = 0)$}; % (2,2)

\node[align=center] at (0.5*\smallwidth, \smallheight + 0.5*\bigheight) {$\bar{n}_1 c_2^{\dagger}$ \\
$c_1^{\dagger} \bar{n}_2$ \\
$(L = \frac{1}{2})$}; % (0,1)

\node[align=center] at (\smallwidth + 0.5*\bigwidth, \smallheight + 0.5*\bigheight) { \begin{tabular}{@{}l@{\hspace{0.5em}}l@{}}
    $n_1 \bar{n}_2 + \bar{n}_1 n_2$ & $(L = 1)$ \\
    $n_1 \bar{n}_2 - \bar{n}_1 n_2$ & $(L = 0)$ \\
    \quad \quad $c_1^{\dagger} c_2$             & $(L = 0)$ \\
    \quad \quad $c_1 c_2^{\dagger}$             & $(L = 0)$ \\
  \end{tabular}
}; % center

\node[align=center] at (\smallwidth + \bigwidth + 0.5*\smallwidth, \smallheight + 0.5*\bigheight) {$n_1 c_2$\\
$c_1 n_2$\\
$(L = \frac{1}{2})$}; % (2,1)

\node[align=center] at (0.5*\smallwidth, 0.5*\smallheight) {$c_1^{\dagger} c_2^{\dagger}$ \\
$(L = 0)$}; % (0,0)

\node[align=center] at (\smallwidth + 0.5*\bigwidth, 0.5*\smallheight) {$n_1 c_2^{\dagger} \quad (L = \frac{1}{2})$ \\
$c_1^{\dagger} n_2 \quad (L = \frac{1}{2})$ }; % (1,0)

\node[align=center] at (\smallwidth + \bigwidth + 0.5*\smallwidth, 0.5*\smallheight) {$n_1 n_2$\\
$(L = 1)$}; % (2,0)

\end{tikzpicture}
\caption{Full operator basis for N = 2 sites, organized by charge sectors $(Q_L, Q_R)$ and total spin $L$ within each block.}
\label{fig:grid}
\end{figure}

\section{SU(2) size distribution and angular momentum for local operators}
In this sector, we study the SU(2) size distribution for local operators. Consider an operator acting on $m_c$ sites with $c$ or $c^\dagger$, $m_n$ sites with $2n-I$, and the remaining $N-m_c-m_n$ sites with the identity operator. Under the mapping between operators and spin configurations introduced in the main text, the local identity operator and $2n-I$ map to $\ket\uparrow + \ket{\downarrow}$ and $\ket\uparrow - \ket{\downarrow}$, which are the eigenstates of $\sigma^x$, denoted as $\ket{\rightarrow}$ and $\ket{\leftarrow}$, respectively. The local creation and annihilation operators map to a spin singlet, denoted as $\ket{0}$, and do not contribute to the angular momentum. As a result, such an operator maps to the following spin configuration,
\begin{equation}
\label{appeq:productspin}
    \ket{\overbrace{0\cdots0}^{m_c} \overbrace{ \leftarrow\cdots \leftarrow}^{m_n} \overbrace{\rightarrow \cdots \rightarrow}^{N-m_c-m_n}}.
\end{equation}
Understanding the SU(2) size distribution of the operator requires expanding this product spin configuration to states with definite total angular momentum. The $m_n$ $\ket{\leftarrow}$ spins and $N-m_c-m_n$ $\ket{\rightarrow}$ spins are fully polarized and have angular momentum $L=m_n/2$ and $L=(N-m_c-m_n)/2$, respectively. Based on the addition of angular momentum, the total spin state has contributions from angular momentum ranging from $(N-m_c)/2$ to $(N-m_c -2 m_n)/2$, and the amplitude of each component can be obtained from the Clebsch-Gordon coefficients. After the decomposition, we get 
\begin{equation}
\ket{\overbrace{0\cdots0}^{m_c} \overbrace{ \leftarrow\cdots \leftarrow}^{m_n} \overbrace{\rightarrow \cdots \rightarrow}^{N-m_c-m_n}} = \sum \limits_{L=|(N-m_c-2m_n)/2|}^{(N-m_c)/2}  \alpha_L \ket{L, L_z=(N-m_c-2m_n)/2}
\end{equation}
where the amplitude is 
\begin{equation}
    \alpha_L =   \sqrt{\frac{ \binom{N-m_c+1}{(N-m_c)/2-L}}{\binom{N-m_c+1}{m_n+1}}\frac{2L+1}{m_n+1}}.
\end{equation}
In the large $N$ limit where $m_c$ and $m_n$ are finite, the expansion is dominated by the minimal $L$ state, whose amplitude is 
\begin{equation}
    \alpha _{L=(N-m_c-2m_n)} = \sqrt{\frac{N-m_c-2m_n+1}{N-m_c-m_n+1}}
\end{equation}
which approaches 1 as $N\rightarrow \infty$. This indicates that the product spin configuration in Eq.~\eqref{appeq:productspin}, although in general has different angular momentum components, becomes an eigenstate of the total angular momentum operator with $L = (N-m_c-2m_n)/2$ in the large $N$ limit. If we define the size $s$ as the difference between the maximal allowed angular momentum $N/2$ and the angular momentum of the operator, then $s = m_c/2+ m_n$. This is equivalent to assigning weight $\{0, 1, 1, 2\}$ to the local operator set $\{I, c, c^\dagger, 2n-I\}_i$ and counting the total weight. This demonstrates that the SU(2) operator size reduces to the definition in previous work for local operators in the large $N$ limit.

\section{Complex Brownian SYK model}
In this section, we provide a detailed derivation of the exact classical master equation from the Brownian SYK model. The derivation, based on emergent SU(4) algebra and permutation symmetry after the disorder average, is based on our previous work. The Hamiltonian of the complex Brownian SYK model is a sum of random time-dependent four-fermion couplings:
\begin{equation}\label{Hamiltonian}
H(t) = \sum_{i,j,k,l} J_{ijkl}(t) c_i^\dagger c_j^\dagger c_k c_l + h.c.    
\end{equation}
where $J_{ijkl}(t)$ are independent complex Gaussian random variables that are uncorrelated in time
\begin{equation}
J_{ijkl}(t) J^*_{i'j'k'l'}(t') = \frac{J}{2N^3}\delta(t-t') \delta_{ii'}\delta_{jj'}\delta_{kk'}\delta_{ll'}.    
\end{equation}
For later convenience, we explicitly enforce the Hermiticity of the Hamiltonian by including the Hermitian conjugate, rather than imposing additional constraints on the couplings $J_{ijkl}$. Moreover, the indices $i, j, k, l$ are allowed to range freely from 1 to $N$. In the summation, multiple terms may correspond to the same operator due to the anticommutation relations of the fermionic operators; however, they do not cancel because the $J_{ijkl}$ are treated as independent variables. Overall, the Hamiltonian contains $N^2(N-1)^2$ terms, with contributions involving $i = j$ or $k = l$ vanishing.

The operator size distribution of an Heisenberg operator $O(t)$ in a basis $S$ is
\begin{equation}
P(\mathcal{S}, t) =\tr \left(O^\dagger (t) \S \right) \tr\left( O(t) \S^\dagger\right).    
\end{equation}
Studying the time evolution of the operator size distribution average over the couplings $J$ requires a random average of four replicas of the time evolution operator. In the replica notation, the distribution is
\begin{equation}
P(\mathcal{S},t) = \bra{\S \otimes S^\dagger} U^a(t)\otimes U^{*,b}(t) \otimes U^{c}(t) \otimes U^{*,d}(t) \ket{O \otimes  O^\dagger}.    
\end{equation}
Because the couplings are uncorrelated in time, one can Trotterize the unitary time evolution and perform the disorder average at individual infinitesimal time slices. At each time slice, we get
\begin{equation}
\mathbb{U}(t+dt, t) =\overline {U^a(t+dt, t)\otimes U^{*,b}(t+dt, t) \otimes U^{c}(t+dt, t) \otimes U^{*,d}(t+dt,t)} = \exp(-\mathbb{H}dt)
\end{equation}
The operator $\mathbb{H}$ is a positive time-independent Hermitian operator acting on all four replicas, dubbed as the emergent Hamiltonian. It reads
\begin{equation}
\begin{aligned}
    &\mathbb{H} = \frac{1}{2}\left(\sum_{A} W_{A} W_{A}^\dagger + W_{A}^\dagger W_{A} \right), \ \ W_{A} =  W_{A}^a - W_{A}^{b,T} + W_{A}^c - W_{A}^{d,T}.
\end{aligned}
\end{equation}
where $A$ is a group index $(i,j,k,l)$ and the operator $W_A^\alpha = c^\dagger_{\alpha,i}c^\dagger_{\alpha,j}c_{\alpha,k}c_{\alpha,l}$ with $\alpha =a,b,c,d$ labels each replica. 

Because $\mathbb{H}$ is independent of $t$. The averaged unitary time evolution on four replicas becomes the imaginary time evolution with respect to $\mathbb{H}$,
\begin{equation}
 \mathbb{U}(t) =  \overline{U^a(t)\otimes U^{*,b}(t) \otimes U^{c}(t) \otimes U^{*,d}(t) }= \exp(-\mathbb{H} t).
\end{equation}

\subsection{Conserved quantities}
The emergent Hamiltonian is a many-body Hamiltonian acting on four copies of the original system, with a local Hilbert space dimension of $16$ and a total Hilbert space dimension of $16^N$. It is invariant under permutations of the site indices. The permutation symmetry is not present in each individual random realization but emerges after performing the disorder average. Consequently, the permutation acts collectively across all replicas, rather than within each replica separately. Restricting to the permutation-invariant subspace reduces the Hilbert space dimension to $\binom{N+15}{15} \sim N^{15}$, representing a substantial reduction from $16^N$, though it remains challenging to work with. Nevertheless, this shows how a random average can significantly reduce the computational complexity of the Brownian SYK model.

Remarkably, by identifying other conserved quantities of the emergent Hamiltonian, we can further show that the Hilbert space of $\H$ fragments into exponentially many sectors, each with a dimension that scales predominantly linearly with $N$.  

We now simplify the emergent Hamiltonian to make its symmetry manifest. We notice that $W_{ij k l} = W^\dagger_{lkji}$. 
Then the two summations in $\mathbb{H}$ are identical and can be combined,
\begin{equation}
    \mathbb{H} = \sum_{A} W_{A} W_{A}^\dagger.
\end{equation}
When one expands $W$, there are 16 terms, including 4 intra-replica terms and 12 inter-replica terms. These terms are quite complicated. But due to the permutation symmetry, these terms only depend on the sum of the fermionic bilinear $S^{\alpha\beta}$ across replicas defined below:
$$
S^{\alpha \beta } = \sum_i \psi_{\alpha, i}^\dagger \psi_{\beta,i}, \quad \psi_{a,i} = c_{a,i}, \ \psi_{b,i} = \mathcal{F}_a c_{b,i}^T, \ \psi_{c,i} = \mathcal{F}_{a}\mathcal{F}_{b}c_{c,i}, \ \psi_{d,i} = \mathcal{F}_{a}\mathcal{F}_{b}\mathcal{F}_c c_{d,i}^T,
$$
where $\mathcal{F}_{\alpha} = \prod(1-2 n_{\alpha,i})$ denotes fermionic parity operator for replica $\alpha$. They are introduced so that $\psi_{\alpha,i}$ anticommutes not only within each replica but also across replicas, thereby preserving the usual fermionic algebra. As a result, the operators $S^{\alpha\beta}$ obey the standard SU(4) commutation relations.
$$
[S^{\alpha\beta}, S^{\gamma\delta}] = \delta_{\beta\gamma} S^{\alpha \delta} - \delta_{\alpha\delta} S^{\beta\gamma}.
$$
The emergent Hamiltonian, which depends only on these operators, does not possess full SU(4) symmetry but commutes with the Casimir operator of the SU(4) algebra. This means that we can further decompose the permutation-invariant subspace to the irreducible representation of the SU(4) algebra, which is labeled by the Young tableau. Furthermore, the original Hamiltonian conserves the total charge and thus the emergent Hamiltonian conserves the total charge in each replica separately. These total charges are given by $S^{\alpha\alpha}$ which corresponds to the weights of the irreducible representation SU(4). 
As a result, the total $16^N$ dimensional Hilbert space is block diagonalized by the irreps of the permutation, the irreps of the SU(4), and the total charge in each replica. For later convenience, we define the charge of each replica as
\begin{equation}
q_a = S^{aa}, \ q_b = (N-S^{bb}), \ q_c = S^{cc}, \ q_d = (N-S^{dd}).
\end{equation}
As we show later, $q_a$ and $q_b$ directly probe the left and right charge of an operator in a system with charge conservation and is consistent with the notation of the main text.

Now we provide the explicit form of the 16 terms and thus the Hamiltonian below. We first consider the 4 intra-replica term
\begin{equation}
\label{appeq:H_piece}
\begin{aligned}
    H_{aa} = \sum_A W_{A,a} W_{A,a}^\dagger = \sum_{ijkl} \psi_{a,i}^\dagger \psi_{a,j}^\dagger \psi_{a,k} \psi_{a,l} \psi_{a,l}^\dagger \psi_{a,k}^\dagger \psi_{a,j} \psi
    _{a,i} = q^a (q^a-1) (N-q^a+2) (N-q^a+1)\\
    H_{bb} = \sum_A W_{A,b}^T W_{A,b}^{\dagger,T} = \sum_{ijkl} \psi_{a,i} \psi_{a,j} \psi_{a,k}^\dagger \psi_{a,l}^\dagger \psi_{a,l} \psi_{a,k} \psi_{a,j}^\dagger \psi_{a,i}^\dagger = q^b (q^b-1) (N-q^b+2) (N-q^b+1)\\
\end{aligned}
\end{equation}
The term $H_{cc}$ and $H_{dd}$ take the same form as well. 

The inter-replica term $H_{\alpha \beta}$ is symmetric under exchange of the replica indices because $W_{\alpha,A}$ and $W_{\beta, A}$ commute and $W_A$ and $W_A^\dagger$ are related by permutation of the site indices. Furthermore, $H_{ab}$, $H_{ad}$, $H_{cb}$ and $H_{cd}$ take the same form after relabeling. Therefore, we only need to compute three terms
\begin{equation}
\begin{aligned}
    &\H_{ac} = \sum_A W_{A,a} W_{A,c}^\dagger = \sum_{ijkl} \psi_{a,i}^\dagger \psi_{a,j}^\dagger \psi_{a,k} \psi_{a,l} \psi_{c,l}^\dagger \psi_{c,k}^\dagger \psi_{c,j} \psi_{c,i} = (S^{ac}S^{ca})^2 -S^{ac}S^{ca}(3q^a+q^c-2)+2q^a(q^a-1), \\
    &\H_{bd} = \sum_A W_{A,b}^T W_{A,b}^{\dagger,T} = \sum_{ijkl} \psi_{b,i} \psi_{b,j} \psi_{b,k}^\dagger \psi_{b,l}^\dagger \psi_{d,l} \psi_{d,k} \psi_{d,j}^\dagger \psi_{d,i}^\dagger = (S^{db}S^{bd})^2 -S^{db}S^{bd}(3q^b+q^d-2)+2q^b(q^b-1), \\
    &\H_{ab} = \sum_A W_{A,a} W_{A,b}^{\dagger,T} = \sum_{ijkl} \psi_{a,i}^\dagger \psi_{a,j}^\dagger \psi_{a,k} \psi_{a,l} \psi_{b,i} \psi_{b,j} \psi_{b,k}^\dagger \psi_{b,l}^\dagger = (S^{ab}S^{ba})^2 + S^{ab}S^{ba}(q^a + q^b -N -2).
\end{aligned}
\end{equation}
Note that $H^{bd}$ can be obtained by relabeling $H^{ac}$ followed by a particle-hole transformation on $b$ and $d$ replicas.
Now the emergent Hamiltonian is expressed a function of $S^{\alpha\beta}S^{\beta\alpha}$, which explicitly conserves the charge $q^{a}\sim q^d$.

\subsection{Operator size basis and SU(4) algebra}
The operator size distribution averaged over the random coupling becomes the overlap of two states under imaginary time evolution
\begin{equation}
P(\S,t) = \bra{O\otimes O^\dagger} \exp(-\H t) \ket{\S\otimes \S^\dagger}.
\end{equation}
In the main text, we propose an operator basis $\S_i(L, L_z,C)$ that is labeled by the operator's angular momentum $L$ and charge. Here $L_z$ and $C$ are related to the left charge and the right charge as
\begin{equation}
    C = q^a - q^b, \quad L_z = (q^a + q^b-N)/2.
\end{equation}
We further aggregate the operator size distribution based on the angular momentum,
\begin{equation}
    P^{L_z, C}(L,t)  = \bra{O\otimes O^\dagger} \exp(-\H t) \sqrt{D^{L_z, C}(L)}\ket{L, L_z,C}, \ket{L,Lz,C} = \frac{ \sum_i \ket{\S_i(L,L_z,C)\otimes \S_i^\dagger (L,L_z,C)}
    }{\sqrt{D^{L_z, C}(L)}} .
\end{equation}
where $D^{L_z, C}(L)$ denotes the number of operators with a fixed $L$, $L_z$ and $C$. 

To take advantage of the SU(4) structure of the emergent Hamiltonian, one can decompose the state $\ket{L,L_z,C}$ into various irreps of SU(4). The emergent Hamiltonian does not mix irreps. Furthermore, the Hamiltonian conserves the charge on each replica, resulting in a much smaller Hilbert space to work with. 
By construction, the state $\ket{L, L_z, C}$ is invariant under permutation and falls into a single SU(4) irrep $(0, N, 0)$ given by the Young tableau,
\begin{equation}
\begin{tikzpicture}[scale=0.8, every node/.style={scale=0.8}]
  \def\boxw{0.8}
  \def\boxh{0.8}

  % First row: 3 boxes, dots, last box
  \foreach \i in {0,1,2} {
    \draw (\i*\boxw, 0) rectangle ++(\boxw, \boxh);
  }
  \node at (3.6*\boxw, 0.5*\boxh) {$\cdots$};
  \draw (4.6*\boxw, 0) rectangle ++(\boxw, \boxh);

  % Second row: 3 boxes, dots, last box
  \foreach \i in {0,1,2} {
    \draw (\i*\boxw, -\boxh) rectangle ++(\boxw, \boxh);
  }
  \node at (3.6*\boxw, -0.5*\boxh) {$\cdots$};
  \draw (4.6*\boxw, -\boxh) rectangle ++(\boxw, \boxh);

  % Bracket with label
  \draw[decorate,decoration={brace,mirror,amplitude=6pt}]
    (0,-1.5*\boxh) -- (5.4*\boxw,-1.5*\boxh)
    node[midway,below=6pt] {$N$ boxes};
\end{tikzpicture}
\end{equation}
The dimension of this irrep is $(N+3)(N+2)^2(N+1)/12$. Each state in this irrep is labeled by the Gelfand-Tsetlin pattern~(see a detailed discussion in~\cite{alex2011numerical})
\begin{equation}
    \begin{pmatrix}
    N & & N & & 0 & & 0\\
    & N & & w &&  0 & & \\
    & & x & & y & & \\
    & & & z & & &
    \end{pmatrix}
\end{equation}
where $w$, $x$, $y$ and $z$ satisfy the constraint that $y\leq w\leq x\leq N$ and $y\leq z\leq x\leq N$. Among these integers, $w$, $x+y$ and $z$ are completely fixed by the weights of SU(4) and cannot change under the imaginary time evolution of the emergent Hamiltonian. Therefore, we can work with a much smaller Hilbert space spanned by states labeled by $x-y$ which is bounded by $N$. 

For the state $\ket{L, L_z, C}$ we consider, since it is a superposition of $\S\otimes \S^\dagger$, we have $q^a=q^d$ and $q^b=q^c$. The state corresponds to the Gelfand-Tsetlin below,
\begin{equation}
    \begin{pmatrix}
    N & & N & & 0 & & 0\\
    & N & & L_z + C/2 -N/2  &&  0 & & \\
    & & N/2+C/2+L/2 & & N/2+C/2-L/2 & & \\
    & & & L_z + C/2 + N/2 & & &
    \end{pmatrix}
\end{equation}
in which the constraint becomes $|L_z|\leq L \leq (N-|C|)/2$. The emergent Hamiltonian is closed under this basis. 
Insert the resolution identity and we get a master equation governing the time evolution of the operator probability distribution
\begin{equation}
    \partial_t P^{L_z, C}(L) = \sum_{L'} A^{L_z, C}_{LL'} P^{L_z, C}(L'), \quad A_{LL'}^{L_z,C} = \sqrt{\frac{D^{L_z,C}(L)}{D^{L_z,C}(L')}} \bra{L,L_z, C} \H \ket{L', L_z,C} 
\end{equation}

Conveniently, the matrix elements of $S^{\alpha\beta}$ in the Gelfand-Tsetlin basis are explicitlyd given in~\cite{alex2011numerical}. From these operators, we can assemble each piece in Eq.~\eqref{appeq:H_piece} to get the matrix element $A_{LL'}^{L_z,C}$, which is presented in the main text.

\section{Solving the master equation}
In this section, we solve the master equation in the large-$N$ limit to obtain the time-dependent probability distribution of operator sizes. We begin by considering an initial operator that is local, such as $c$ or $c^\dagger$ on a single site and the identity on all other sites. Such operators have a fixed value of $C$ but generally distribute weight across multiple $L_z$ sectors. We therefore work in a scaling limit where $l_z = L_z/N$ remains fixed as $N \to \infty$.
We define the operator size as $s = N/2 - L$, which increases over time as $L$ decreases from its maximal value $(N - C)/2$. For a simple initial operator with a small value of $C$ compared to $N$, the size remains finite in the early time. In other words, the initial size is $s_0 = \mathcal{O}(1)$ and does not scale with the system size.
In this limit, the transition rates reduce to the following simple expressions: 
\begin{equation}
    \lim_{N \to \infty} \gamma^-_s = (1 - 4l_z^2)\, s, \quad
    \lim_{N \to \infty} \gamma^+_s = 0.
\end{equation}
As a result, the master equation reduces to a simple first-order recurrence:
\begin{equation}\label{simple_master_eq}
    \partial_t P(s,t) = \lambda \left[(s - 1)P(s - 1,t) - sP(s,t)\right], \quad \lambda = 1 - 4l_z^2.
\end{equation}
This equation can be written in matrix form, where the corresponding transition matrix acting on $P(s,t)$ is lower triangular with eigenvalues given by the non-negative integers $k$. The associated left and right eigenvectors are:
\begin{equation}
\begin{aligned}
    v_k(s) &= \frac{(-1)^{s - k} \, \Gamma(k)}{\Gamma(1 + k - s) \, \Gamma(s)}, \\
    u_k(s) &= \frac{\Gamma(s)}{\Gamma(1 - k + s) \, \Gamma(k)}.
\end{aligned}
\end{equation}
Orthogonality between left and right eigenvectors with different eigenvalues can be explicitly verified:
\begin{equation}
\begin{split}
    \sum_{s=0}^\infty v_{k'}(s) u_{k}(s) &= 
    \frac{\Gamma(k')}{\Gamma(k) \Gamma((k' - k) + 1)} \sum_{l=0}^{(k' - k)} (-1)^l \binom{(k' - k)}{l} = \delta_{k',k}.
\end{split}
\end{equation}
Given an initial operator of fixed size $s_0$, the early-time probability distribution takes the form:
\begin{equation}
\begin{split}
    P(s,t) &= \sum_{k \geq s_0} v_k(s_0) u_k(s) e^{-\lambda k t} = \frac{e^{-s_0 \lambda t} (1 - e^{-t\lambda})^{s-s_0} \Gamma(s)}
    {\Gamma(1 + s-s_0) \Gamma(s_0)}.
\end{split}
\end{equation}
For $s < s_0$ the probability distribution vanishes. This follows directly from the expression above, which is defined only for $s \ge s_0$ since $\Gamma(1 + s - s_0)$ develops poles at non-positive integers. The master equation is fully consistent with this restriction, as its dynamics permit only upward transitions in $s$ and therefore cannot populate sizes below the initial value.
In the regime where the operator size becomes extensive, it is convenient to introduce the continuum variable $x = s/N$. Taking the large-$N$ limit with $x$ held fixed, the probability distribution takes the form
\begin{equation}
    P(x,t) = \frac{e^{-s_0 \lambda t} (1 - e^{-t\lambda})^{Nx} (Nx)^{s_0}}
    {x \Gamma(s_0) (1 - e^{-t\lambda})^{s_0}}.
\end{equation}
To proceed, we introduce the scrambling time:
\begin{equation}
    t^* = \frac{\log N}{\lambda}.
\end{equation}
At \( t^* \), the probability distribution has a continuum limit which takes the form of a chi-squared distribution:
\begin{equation}
    p(x,t^*) = f_{s_0}(x) = \frac{1}{\Gamma(s_0)} e^{-x} x^{s_0 - 1}.
\end{equation}
Since our analytical solution is valid for early times, we now consider the probability distribution at \( t^* + \tau \), where \( \tau \to -\infty \):
\begin{equation}\label{initial}
\begin{split}
    p(x,t^*+\tau) &= \frac{e^{-\lambda \tau}}{\Gamma(s_0)} e^{-e^{-\lambda \tau} x} (e^{-\lambda \tau} x)^{s_0 - 1} = e^{-\lambda \tau} f_{s_0}(e^{-\lambda \tau} x).
\end{split}
\end{equation}
For $s_0 = 1/2$, the probability distribution diverges at small $x$ due to the singular behavior of $f_{s_0}(x)$ in this limit. We confirm this divergence by numerically solving the master equation for a simple operator with $C = 1$ (as shown in the main text), which corresponds to $s_0 = 1/2$. This singularity persists into the late-time regime and disappears for larger values of $C$, as expected. The distribution in Eq.~\eqref{initial} will serve as the initial condition for extending the solution beyond the early-time regime.

\subsection{Beyond Early Time}
To obtain the full-time evolution of the probability distribution, all transition processes must be considered. For this purpose, we rewrite the dynamics in terms of the continuum variable $l = L/N$ employed in the main text. The operator-size fraction $x = s/N$ is related to $l$ through $x = \tfrac{1}{2} - l$, so expressing the early-time distribution in terms of $l$ leads to a more compact form for the full-time behavior. In this continuum limit, the master equation reduces to the Fokker–Planck equation:
\begin{equation}
    \partial_t p(l,t) = -\partial_l \big( v(l) p(l,t) \big) + \frac{1}{N} \partial_l^2 \big( d(l) p(l,t) \big).
\end{equation}
where \( v(l) \) and \( d(l) \) are the drift and diffusion coefficients, respectively. In the large-$N$ limit, the diffusion term becomes negligible, and the dynamics are governed primarily by deterministic drift. The resulting equation can be solved using the method of characteristics. By applying the Green’s function, the time evolution of the distribution is given by:
\begin{equation}
\begin{aligned}
    p(l,t) &= \int dl_0 \, \delta \Big( l - \bar{l} \big( t - (t^* +\tau) + \bar{l}^{-1}(l_0) \big) \Big) p(l_0, t^* + \tau).
\end{aligned}
\end{equation}
where $p(l_0, t^*+\tau)$ is the early-time distribution in Eq.~\eqref{initial}, written in terms of $l_0 = \tfrac{1}{2} - x_0$, and $\bar{l}(t)$ is the solution of the characteristic equation:
\begin{equation}
    \frac{d \bar{l}}{dt} = v(\bar{l})
\end{equation}
and $\bar{l}^{-1}$ is the inverse function of $\bar{l}$. The resulting full-time distribution retains the chi-squared form under a change of variables:
\begin{equation}\label{full_time}
    p(l,t) = \frac{d \xi}{dl} f_{s_0}(\xi) = \frac{d\xi}{dl} \frac{1}{\Gamma(s_0)} e^{-\xi} \xi^{s_0 - 1}.
\end{equation}
The transformed variable $\xi$ is defined as
\begin{equation}
    \xi = e^{-\lambda \tau} \bar{l} \big( \bar{l}^{-1}(l) + t^* + \tau - t \big).
\end{equation}
which ensures that the distribution reduces to the initial condition at $t = t^* + \tau$.
The drift velocity $v(l)$ appearing in the characteristic equation is
\begin{equation}
    v(l) = \frac{(l^2 - l_z^2)\big(4l^2 -1 )}{2l}.
\end{equation}
In the limit where \( \tau \ll 0 \) with \( t - t^*\) fixed, the variable $\xi$ simplifies to:
\begin{equation}
     \xi = \frac{e^{(t^* - t)\lambda} \, \lambda(1 - 4l^2)}{16\left(l^2 -l_z^2\right)}.
\end{equation}
Substituting this expression into Eq.~\eqref{full_time} yields the
full-time probability distribution. In the main text, we verify this
result by numerically solving the full master equation for simple
initial operators with small charge $C$. As the system size $N$ is
increased, the numerical distributions converge toward the
large-$N$ analytic expression Eq.~\eqref{full_time}, showing excellent
agreement in the regime where finite-size effects are suppressed.

We now turn to the regime, where the charge scales extensively, $C=\mathcal{O}(N)$, so
that $c=C/N$ remains finite in the large N limit. In this regime, the quantities $L$, $L_z$, and $C$ all scale extensively with system size, and the system rapidly loses memory of its initial conditions. As a result, the time evolution of the operator size can be obtained directly by solving the characteristic equation. 
The drift velocity in this case takes the form:
\begin{equation}
    v(l,l_z,c) = \frac{(l^2 - l_z^2)(4l^2 - (1-c^2))}{2l}.
\end{equation}
The solution to the drift equation follows as:
\begin{equation}
    l^2(t) = \frac{\lambda}{4+4e^{\lambda(t-t_0)}}+l_z^2, \quad \lambda = 1-c^2 - 4l_z^2.
\end{equation}
Here, \( t_0 \) depends on the initial condition. 
% At \( t = 0 \), we consider the largest possible value of \( l \), given by \( l = (1 - \lvert c \rvert)/2 \). Under this assumption, \( t_0 \) is determined as:
% \begin{equation}
%     t_0 = \frac{1}{\lambda} \log \left(\frac{-4l_z^2+(\lvert c \rvert-1)^2}{8\lvert c \rvert(1-\lvert c \rvert)} \right).
% \end{equation}
In this regime where $c$ remains finite, the probability distribution
collapses onto a sharply peaked profile centered around the
deterministic trajectory $l(t)$ that flows from its maximal initial
value toward $l_z$. This behavior is confirmed numerically in the main
text, where simulations of the full master equation show the
distribution narrowing and tracking the characteristic solution as
$N$ increases.

\begin{figure}
    \centering
    \includegraphics[width=0.32\linewidth]{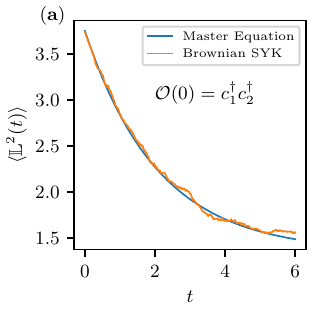}\hfill
    \includegraphics[width=0.32\linewidth]{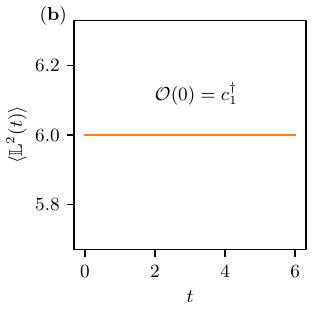}\hfill
    \includegraphics[width=0.32\linewidth]{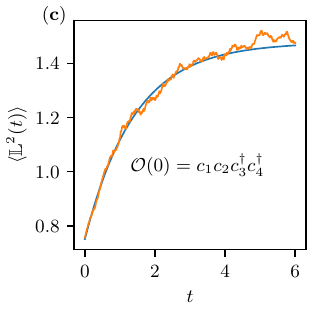}
    \caption{Comparison of operator time evolution under the complex Brownian SYK Hamiltonian and the master equation for $N = 5$. Disorder averaging is performed over 15 independent realizations of complex random couplings.}
    \label{fig:benchmark}
\end{figure}

\section{Benchmark with the noisy Hamiltonian}
In this section, we benchmark the complex Brownian SYK Hamiltonian against the master equation. Using the superoperator formalism introduced in the main text, the time evolution of the squared angular momentum $\braket{\L^2}$ can be computed as:
\begin{equation}
\begin{aligned}  
    \braket{\L^2(t)} = \sum_{ij} \text{tr} \left[ \mathcal{O}^\dagger(t) \, c_i^\dagger c_j \, \mathcal{O}(t) \, c_j^\dagger c_i \right] + \braket{L_z (L_z - 1)}.
\end{aligned}
\end{equation}
Here, the first term is an out-of-time-order correlator (OTOC), while the second term remains constant in time.
As shown in Fig.~\ref{fig:benchmark}, we find excellent agreement between the stochastic time evolution under the Brownian SYK Hamiltonian~\eqref{Hamiltonian} and the master equation simulation across different charge sectors for a system of size $N = 5$.
In Fig.~\ref{fig:benchmark}(a), we consider the initial operator $\mathcal{O}(0) = c_1^\dagger c_2^\dagger$, with identity operators on the remaining sites. Since this operator does not belong to a well-defined charge sector, we project it onto the sector with $Q_L = 3$ and $Q_R = 1$, yielding the operator $c_1^\dagger c_2^\dagger (n_3 \bar{n}_4 \bar{n}_5 + \bar{n}_3 n_4 \bar n _5 + \bar n_3 \bar n_4 n_5)$. The corresponding angular momentum starts at its maximum value, $L=3/2$, and decays toward the steady-state value.
In Fig.~\ref{fig:benchmark}(b), the initial operator $\mathcal{O}(0) = c_1^\dagger$ is projected into the charge sector $Q_L = 1$, $Q_R = 0$. The resulting operator is $c_1^\dagger \bar{n}_2 \bar{n}_3 \bar{n}_4 \bar{n}_5$  with total angular momentum $L = 2$.  In this case, the angular momentum remains constant in time, as the corresponding operator is the only operator in this sector.
In Fig.~\ref{fig:benchmark}(c), we consider the operator
$\mathcal{O}(0) = c_1 c_2 c_3^\dagger c_4^\dagger$ projected to the
sector $Q_L = Q_R = 2$. In this special case, the projected operator is $c_1 c_2 c_3^\dagger c_4^\dagger \bar n_5$ with minimum allowed angular momentum $L = 1/2$ in this sector. As a result, the operator evolves
toward configurations with larger $L$, eventually
relaxing to the corresponding steady-state value.

At late times, all operator basis states within a fixed charge sector become equally probable. Using the degeneracy factor $D^{L_z, C}(L)$ derived in Eq.~\eqref{eq:degeneracy}, we obtain a closed-form expression for the late-time average:
\begin{equation}\label{eq:late_time}
\lim_{t \to \infty} \braket{\mathbb{L}^2(t)} = \frac{\sum_L L(L+1) D^{L_z, C}(L)}{\mathcal{N}(L_z,C)} = L_z^2 + \frac{N^2 - C^2 + 4L_z^2}{4N},
\end{equation}
where $\mathcal{N}(L_z, C)$ is the total number of operator basis states in the given charge sector:
\begin{equation}
\mathcal{N}(L_z, C) = \sum_{L = |L_z|}^{(N-C)/2} D^{L_z, C}(L) =\binom{N}{\frac{N-|C|}{2} - |L_z|} \binom{N}{\frac{N+|C|}{2} - |L_z|}.
\end{equation}
In diagonal sectors ($C = 0$), such as the example shown in Fig.~\ref{fig:benchmark}(c), Eq.~\eqref{eq:late_time} slightly overestimates the steady-state value in finite-$N$ systems due to the presence of the identity operator in the late-time ensemble. However, for any nontrivial initial operator in these sectors, the identity operator remains inaccessible at all times. Accordingly, it should be excluded from the ensemble average. The corrected late-time value is therefore:
\begin{equation}
\lim_{t \to \infty} \braket{\mathbb{L}^2(t)} = \frac{\displaystyle\sum_{L = |L_z|}^{(N-C)/2} L(L+1) D^{L_z, C}(L) - \frac{N}{2}(\frac{N}{2}+1)\delta_{C,0}}{\mathcal{N} (L_z, C)- \delta_{C,0}}.
\end{equation}
The resulting correction vanishes exponentially with $N$ and becomes negligible in the large-$N$ limit. In the late time, the scaled angular momentum $\braket{l^2}$ approaches $l_z^2$ as $N \to \infty$.

\end{document}